\documentclass[lettersize, journal]{IEEEtran}
\usepackage{amsmath,amsfonts}
\usepackage{array}
\usepackage[caption=false,font=normalsize,labelfont=sf,textfont=sf]{subfig}
\usepackage{textcomp}
\usepackage{stfloats}
\usepackage{url}
\usepackage{verbatim}
\usepackage{graphicx}
\usepackage{cite}
\usepackage{amsmath,amssymb}
\usepackage{algorithm}
\usepackage{algorithmic}
\usepackage{subfig} 
\usepackage{makecell}
\usepackage[top=0.61in, bottom=0.94in, left=0.57in, right=0.57in]{geometry}
\usepackage{multirow} 

\usepackage{amsthm}

\usepackage{graphicx} 
\usepackage{amsmath}
\usepackage{amsmath}
\usepackage{amsmath, amssymb}
\usepackage{xcolor}	
\usepackage{amsmath} 

\usepackage{graphics}
\usepackage{amsmath, amsthm}

\theoremstyle{plain}

\begin{document}

\title{PASS-Assisted RSMA under Imperfect SIC: Joint Antenna Activation and Resource Allocation}

\author{
Saeid Pakravan,
Imene Trigui,
Wessam Ajib,
and Wei-Ping Zhu
\thanks{
S. Pakravan and W. Ajib are with the Department of Computer Science, University of Quebec in Montreal (UQAM), Montreal, QC, Canada
(e-mail: pakravan.saeid@uqam.ca; ajib.wessam@uqam.ca).
}
\thanks{
I. Trigui is with the Department of Applied Sciences, University of Quebec at Chicoutimi (UQAC), Chicoutimi, QC, Canada
(e-mail: itrigui@uqac.ca).
}
\thanks{
W.-P. Zhu is with the Department of Electrical and Computer Engineering, Concordia University, Montreal, QC, Canada
(e-mail: weiping@ece.concordia.ca).
}
}

	\maketitle

\begin{abstract}

The performance of rate-splitting multiple access (RSMA) can be severely affected by imperfect successive interference cancellation (SIC) in practical wireless systems. This paper investigates a downlink pinching antenna system (PASS)-assisted RSMA network under imperfect SIC, where residual common-stream interference is explicitly incorporated into private-stream decoding. To improve user fairness, a max–min rate optimization problem is formulated through the joint design of antenna activation, common-rate allocation, and power allocation. The resulting mixed-integer non-convex problem is addressed using a two-stage framework that combines greedy channel-aware antenna activation with successive convex approximation (SCA)-based resource allocation. Numerical results demonstrate the effectiveness of the proposed framework in improving fairness under imperfect SIC.

\end{abstract}


\section{Introduction}

Future beyond-5G and 6G wireless networks are expected to support massive connectivity and diverse quality-of-service requirements, making efficient interference management a critical challenge. Among emerging multiple-access technologies, rate-splitting multiple access (RSMA) has attracted significant attention due to its ability to partially decode interference while treating the remaining interference as noise, thereby providing a flexible transmission framework that bridges space-division multiple access and non-orthogonal multiple access (NOMA) \cite{10038476}. Owing to its adaptability to heterogeneous channel conditions, RSMA has demonstrated considerable gains in spectral efficiency, reliability, and fairness across a wide range of communication scenarios \cite{9832611}. Nevertheless, the performance of RSMA remains highly dependent on effective channel quality.

To further improve channel conditions, pinching antenna systems (PASSs) have recently emerged as a promising reconfigurable antenna architecture that enables waveguide-based spatial adaptability with a limited number of radio-frequency chains \cite{11169486,10945421, 11481959, 10976621, Pakravan2026PASAirComp}. By selectively activating antenna elements along dielectric waveguides, PASS can dynamically reconfigure the wireless propagation environment and improve effective user channels. This capability makes PASS particularly attractive for RSMA systems, whose performance strongly depends on channel conditions. Consequently, integrating PASS with RSMA offers a promising approach for jointly enhancing channel quality and interference-management capability \cite{11304137,11551681,Wang2026PASSRSMA}.

Despite these advances, existing works generally assume perfect successive interference cancellation (SIC). In practical wireless systems, however, imperfect SIC inevitably arises due to channel estimation errors, hardware impairments, and decoding inaccuracies, resulting in residual interference that degrades private-stream decoding performance \cite{10747269,11223080,11455905}. Although imperfect SIC has been studied in conventional RSMA systems, its impact on PASS-assisted RSMA remains largely unexplored. In particular, antenna activation directly affects user-channel characteristics and thus influences both common-stream decoding and residual-interference propagation. Consequently, resource-allocation strategies developed under perfect SIC assumptions may become suboptimal in practical PASS-assisted RSMA deployments. 

Motivated by this gap, this paper investigates a downlink multi-user PASS-assisted RSMA system under imperfect SIC. Residual common-stream interference is explicitly incorporated into private-stream decoding, and a fairness-oriented max--min rate optimization problem is formulated through the joint design of antenna activation, common-rate allocation, and power allocation. To efficiently solve the resulting mixed-integer non-convex problem, a low-complexity two-stage framework is developed, where greedy channel-aware antenna activation is combined with successive convex approximation (SCA)-based resource allocation. Numerical results demonstrate that the proposed framework significantly improves fairness performance and exhibits enhanced robustness against SIC imperfections compared with benchmark schemes.

\section{System Model and Problem Formulation}

\begin{figure}[t]
\centering
\includegraphics[width=0.37\textwidth,height=3.9cm]{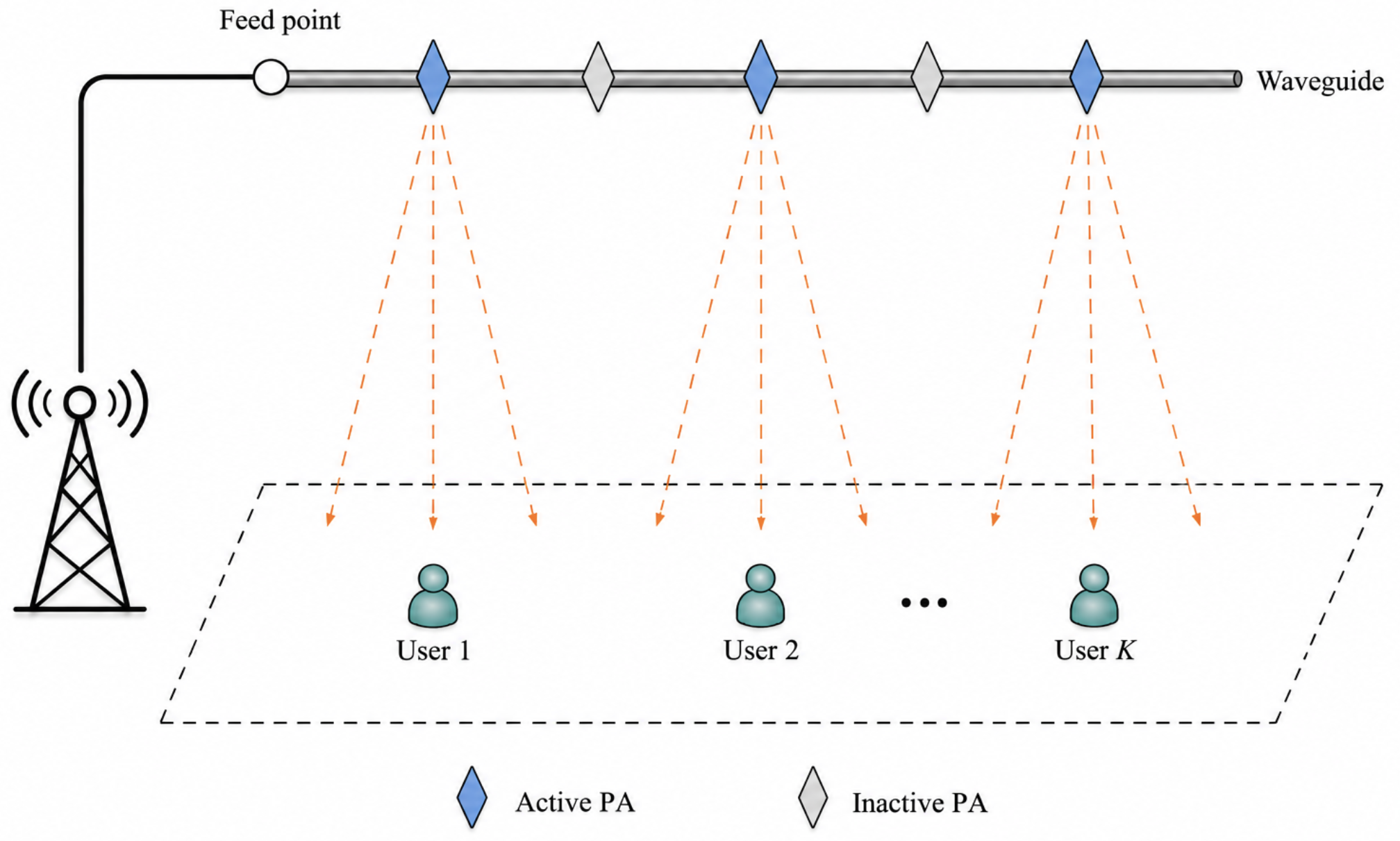}
\caption{Illustration of the proposed PASS-assisted RSMA system model.}
\label{fig_system_model}
\end{figure}

Consider a downlink multi-user PASS-assisted RSMA network, as illustrated in Fig.~\ref{fig_system_model}. A base station (BS) feeds a dielectric waveguide equipped with $N$ preconfigured candidate PA positions to serve $K$ single-antenna users. Let $\mathcal{K}=\{1,\ldots,K\}$ and $\mathcal{N}=\{1,\ldots,N\}$ denote the sets of users and candidate PAs, respectively. Due to hardware and implementation constraints, only $M$ PAs are activated simultaneously, where $M<N$. Let $\alpha_n\in\{0,1\}$ denote the activation status of the $n$-th PA, where $\alpha_n=1$ indicates an active PA and $\alpha_n=0$ otherwise. The position of user $k$ is denoted by
$\boldsymbol{\psi}_k=(x_k,y_k,0)$,
whereas the position of candidate PA $n$ is
$\boldsymbol{\psi}_n^P=(x_n^P,0,d)$,
with $d$ representing the waveguide height. The feed point of the waveguide is denoted by $\boldsymbol{\psi}_0^P$.

\subsection{PASS Channel Model}

Following the commonly adopted PASS channel model for high-frequency LoS-dominant scenarios, only the free-space LoS component is considered \cite{10945421, 6363891}. The signal propagates through the dielectric waveguide and is radiated by the activated PAs. The in-waveguide propagation channel from the feed point to the $n$-th PA is modeled as
\begin{equation}
    g_n=
    e^{-\kappa_{\rm w}l_n}
    e^{-j\frac{2\pi}{\lambda_g}l_n},
\end{equation}
where $\kappa_{\rm w}$ denotes the waveguide attenuation coefficient, $\lambda_g$ is the guided wavelength, and
    $l_n=
    \left\|
    \boldsymbol{\psi}_n^P-\boldsymbol{\psi}_0^P
    \right\|$
is the propagation distance inside the waveguide. The guided wavelength is given by
$\lambda_g=\frac{\lambda}{n_{\rm eff}}$,
where $\lambda$ is the free-space wavelength and $n_{\rm eff}$ denotes the effective refractive index of the dielectric waveguide.

The free-space LoS channel between the $n$-th PA and user $k$ is expressed as
\begin{equation}
    h_{n,k}^{\rm fs}
    =
    \frac{\lambda}{4\pi d_{n,k}}
    e^{-j\frac{2\pi}{\lambda}d_{n,k}},
\end{equation}
where $d_{n,k}
    =
    \left\|
    \boldsymbol{\psi}_n^P-\boldsymbol{\psi}_k
    \right\|$
is the distance between the $n$-th PA and user $k$.

Accordingly, the composite PASS channel from the BS feed point to user $k$ through the $n$-th PA is
\begin{equation}
    h_{n,k}=g_n h_{n,k}^{\rm fs}.
\end{equation}
Following the equal-power radiation model commonly adopted in discrete PASS designs \cite{11169486, 11202577, 11263923}, the transmitted signal is uniformly distributed among the activated PAs. Accordingly, the equivalent BS-user channel is
\begin{equation}
    h_k(\boldsymbol{\alpha})
    =
    \frac{1}{\sqrt{M}}
    \sum_{n=1}^{N}
    \alpha_n h_{n,k},
\end{equation}
where
    $\boldsymbol{\alpha}
    =
    [\alpha_1,\alpha_2,\ldots,\alpha_N]^T$
denotes the PA activation vector. The normalization factor $1/\sqrt{M}$ ensures a constant total radiated power across the activated PAs.

The PA activation constraint is given by
\begin{equation}
    \sum_{n=1}^{N}\alpha_n=M,
    \quad
    \alpha_n\in\{0,1\},
    \quad
    \forall n\in\mathcal{N}.
\end{equation}

\subsection{RSMA Transmission}

In RSMA, the message of each user is divided into common and private parts. The common parts of all users are jointly encoded into a common stream $s_c$, while the private parts are independently encoded into private streams $s_k$, $k\in\mathcal{K}$. The transmitted signal is
\begin{equation}
    x=
    \sqrt{p_c}s_c+
    \sum_{k=1}^{K}\sqrt{p_k}s_k,
\end{equation}
where $p_c$ and $p_k$ denote the transmit powers allocated to the common stream and the private stream of user $k$, respectively.
The transmitted streams satisfy
$\mathbb{E}[|s_c|^2]=1$, $
    \mathbb{E}[|s_k|^2]=1$,
and the BS transmit power constraint is given by
\begin{equation}
    p_c+\sum_{k=1}^{K}p_k
    \le P_{\max},
\end{equation}
where $P_{\max}$ denotes the maximum transmit power.

The received signal at user $k$ is therefore expressed as
\begin{equation}
    y_k=
    h_k(\boldsymbol{\alpha})x+n_k,
\end{equation}
where $n_k\sim\mathcal{CN}(0,\sigma_k^2)$
denotes additive white Gaussian noise.

Following the conventional RSMA decoding procedure, each user first decodes the common stream by treating all private streams as noise \cite{10038476,9832611}. Accordingly, the received signal-to-interference-plus-noise ratio (SINR) for decoding the common stream at user $k$ is 
\begin{equation}
    \gamma_{c,k}
    =
    \frac{
    |h_k(\boldsymbol{\alpha})|^2p_c
    }
    {
    |h_k(\boldsymbol{\alpha})|^2
    \sum_{j=1}^{K}p_j
    +\sigma_k^2
    }.
\end{equation}

Therefore, the achievable common-stream rate at user $k$ is 
\begin{equation}
    R_{c,k}
    =
    \log_2(1+\gamma_{c,k}).
\end{equation}
Since the common stream must be decoded by all users, the achievable common-stream rate is limited by the weakest common-stream decoder:
\begin{equation}
    R_c=
    \min_{k\in\mathcal{K}}
    R_{c,k}.
\end{equation}
Let $C_k$ denote the portion of the common rate allocated to user $k$. The common-rate allocation satisfies
\begin{equation}
    \sum_{k=1}^{K}C_k
    \le
    R_c,
    \quad
    C_k\ge 0,\ \forall k\in\mathcal{K}.
\end{equation}

After decoding the common stream, each user performs SIC before decoding its private stream. Consistent with the conventional single-layer RSMA decoding structure, imperfect SIC is modeled through residual common-stream interference during private-stream decoding. Let $\rho\in[0,1]$ denote the residual interference coefficient, where $\rho=0$ corresponds to perfect SIC. The resulting private-stream SINR is
\begin{equation}
    \gamma_{p,k}
    =
    \frac{
    |h_k(\boldsymbol{\alpha})|^2p_k
    }
    {
    \rho |h_k(\boldsymbol{\alpha})|^2p_c
    +
    |h_k(\boldsymbol{\alpha})|^2
    \sum_{j\neq k}p_j
    +
    \sigma_k^2
    }.
\end{equation}
 The private-stream rate of user $k$ is
\begin{equation}
    R_{p,k}
    =
    \log_2(1+\gamma_{p,k}).
\end{equation}
Hence, the total achievable rate of user $k$ is
\begin{equation}
    R_k=C_k+R_{p,k}.
\end{equation}

\subsection{Problem Formulation}

To improve fairness under imperfect SIC, a max--min rate optimization problem is formulated through the joint design of PA activation, common-rate allocation, and power allocation. Let $\mathbf{p}=[p_c,p_1,\ldots,p_K]^T$ and $\mathbf{C}=[C_1,\ldots,C_K]^T$. The resulting optimization problem is formulated as
\begin{subequations}
\label{prob_original}
\begin{align}
\max_{\boldsymbol{\alpha},\mathbf{C},\mathbf{p}}
\quad &
\min_{k\in\mathcal{K}} R_k
\\
\text{s.t.}\quad
&
\sum_{k=1}^{K} C_k \le R_c,
\label{prob_original_common}
\\
&
C_k \ge 0,
\quad \forall k\in\mathcal{K},
\label{prob_original_C}
\\
&
p_c+\sum_{k=1}^{K}p_k
\le P_{\max},
\label{prob_original_power}
\\
&
p_c,p_k \ge 0,
\quad \forall k\in\mathcal{K},
\label{prob_original_nonnegative}
\\
&
\sum_{n=1}^{N}\alpha_n=M,
\quad
\alpha_n\in\{0,1\},
\quad \forall n\in\mathcal N.
\label{prob_original_activation}
\end{align}
\end{subequations}
Constraint~\eqref{prob_original_common} guarantees feasible common-rate allocation, while \eqref{prob_original_power} imposes the transmit-power budget at the BS. Constraint~\eqref{prob_original_activation} enforces the PASS activation policy. Problem \eqref{prob_original} is a mixed-integer non-convex optimization problem due to the binary PA activation variables and the non-convex coupling among PA activation, power allocation, and achievable-rate expressions.

\section{Proposed Solution}

To solve Problem~\eqref{prob_original}, a two-stage framework is developed. First, a greedy PA activation strategy determines the PASS configuration. Then, for the selected PA set, common-rate and power allocations are optimized using SCA.

\subsection{Greedy PA Activation}

Rate-based PA selection requires solving the resource-allocation problem for every candidate activation pattern, resulting in prohibitive complexity. Therefore, we adopt the weakest effective channel gain as a low-complexity surrogate, since bottleneck users typically determine the max--min rate performance. Although the proposed metric does not directly maximize the minimum user rate, improving the weakest effective channel gain generally enhances the SINRs of bottleneck users and thus promotes max--min fairness.

Let $\mathcal{A}$ denote the set of active PAs. Starting from $\mathcal{A}=\emptyset$, one PA is activated at each iteration. For a candidate PA $n\in\mathcal{N}\setminus\mathcal{A}$, define the temporary active set as
$\mathcal{A}_n = \mathcal{A}\cup\{n\}$.
The corresponding effective channel of user $k$ is evaluated as
\begin{equation}
h_k(\mathcal{A}_n)
=
\frac{1}{\sqrt{|\mathcal{A}_n|}}
\sum_{i\in\mathcal{A}_n}
h_{i,k}.
\end{equation}
The normalization factor $1/\sqrt{|\mathcal{A}_n|}$ ensures a fair comparison among candidate activation sets with different cardinalities during the greedy search. The PA selected at each iteration is determined according to
\begin{equation}
n^\star
=
\arg\max_{n\in\mathcal{N}\setminus\mathcal{A}}
\;
\min_{k\in\mathcal{K}}
|h_k(\mathcal{A}_n)|^2.
\end{equation}
The active set is updated as
$\mathcal{A}
\leftarrow
\mathcal{A}\cup\{n^\star\}$,
and the procedure continues until $|\mathcal{A}|=M$. The resulting activation vector is obtained as
\begin{equation}
\alpha_n=
\begin{cases}
1, & n\in\mathcal{A},\\
0, & \text{otherwise}.
\end{cases}
\end{equation}
For notational convenience, define the effective channel gain for the selected PA configuration as
$H_k
=
|h_k(\boldsymbol{\alpha})|^2,
\forall k\in\mathcal{K}$.

\subsection{SCA-Based Resource Allocation}

For fixed PA activation, the remaining variables are common-rate and power allocations. Introducing an auxiliary variable $\eta$ representing the minimum user rate, Problem~\eqref{prob_original} reduces to the following continuous resource-allocation subproblem:
\begin{subequations}
\label{prob_eta}
\begin{align}
\max_{\mathbf C,\mathbf p,\eta}
\quad &
\eta
\\
\mathrm{s.t.}\quad
&
C_k+R_{p,k}
\ge
\eta,
\quad
\forall k\in\mathcal K,
\\
&
\sum_{k=1}^{K}C_k
\le
R_{c,k},
\quad
\forall k\in\mathcal K,
\\
&
p_c+\sum_{k=1}^{K}p_k
\le
P_{\max},
\\
&
p_c,p_k\ge0,
\quad
\forall k\in\mathcal K,
\\
&
C_k\ge0,
\quad
\forall k\in\mathcal K,
\end{align}
\end{subequations}
where the second constraint is equivalent to
$\sum_{k=1}^{K} C_k \le R_c$ since
$R_c=\min_{k\in\mathcal K} R_{c,k}$.

Problem~\eqref{prob_eta} remains non-convex due to the coupled SINR expressions. To facilitate tractable optimization, the private-stream rate is rewritten as
{\small
\begin{equation}
\begin{aligned}
R_{p,k}
=
&
\log_2
\left(
\rho H_k p_c
+
H_k\sum_{j=1}^{K}p_j
+
\sigma_k^2
\right)
\\
&
-
\log_2
\left(
\rho H_k p_c
+
H_k\sum_{\substack{j=1, j\neq k}}^{K}p_j
+
\sigma_k^2
\right),
\end{aligned}
\end{equation}
}
while the common-stream rate can be expressed as
\begin{equation}
\small
R_{c,k}
=
\log_2\!\left(H_k p_c + H_k\sum_{j=1}^{K}p_j + \sigma_k^2\right)
-
\log_2\!\left(H_k\sum_{j=1}^{K}p_j + \sigma_k^2\right).
\end{equation}
Both achievable-rate expressions admit a difference-of-concave representation. Since the logarithmic function is concave, its first-order Taylor expansion provides a global upper bound. Specifically, for a generic concave function $g(\mathbf p)$,
\begin{equation}
g(\mathbf p)
\le
g(\mathbf p^{(t)})
+
\nabla g(\mathbf p^{(t)})^T
(\mathbf p-\mathbf p^{(t)}),
\end{equation}
where $\mathbf p^{(t)}$ denotes the operating point at iteration $t$.

For the private-stream rate, define
\begin{equation}
B_{p,k}
=
\rho H_k p_c
+
H_k\sum_{\substack{j=1, j\neq k}}^{K}p_j
+
\sigma_k^2 .
\end{equation}
Applying the first-order Taylor approximation to $\log_2(B_{p,k})$ at $\mathbf p^{(t)}$ yields
\begin{equation}
\widehat B_{p,k}^{(t)}
=
\log_2(B_{p,k}^{(t)})
+
\frac{B_{p,k}-B_{p,k}^{(t)}}
{B_{p,k}^{(t)}\ln 2}.
\end{equation}
Accordingly, a concave lower bound of $R_{p,k}$ is obtained as
\begin{equation}
\small
\widetilde R_{p,k}^{(t)}
=
\log_2
\left(
\rho H_k p_c
+
H_k\sum_{j=1}^{K}p_j
+
\sigma_k^2
\right)-
\widehat B_{p,k}^{(t)}.
\end{equation}
Similarly, a concave lower bound $\widetilde R_{c,k}^{(t)}$ of $R_{c,k}$ is obtained by first-order approximation of the second logarithmic term. Consequently, the following convex approximation is solved at each iteration:
\begin{subequations}
\label{prob_sca}
\begin{align}
\max_{\mathbf C,\mathbf p,\eta}
\quad &
\eta
\\
\mathrm{s.t.}\quad
&
C_k+\widetilde R_{p,k}^{(t)}
\ge
\eta,
\quad
\forall k\in\mathcal K,
\\
&
\sum_{k=1}^{K}C_k
\le
\widetilde R_{c,k}^{(t)},
\quad
\forall k\in\mathcal K,
\\
&
p_c+\sum_{k=1}^{K}p_k
\le
P_{\max},
\\
&
p_c,p_k\ge0,
\quad
\forall k\in\mathcal K,
\\
&
C_k\ge0,
\quad
\forall k\in\mathcal K.
\end{align}
\end{subequations}
Since the resulting rate lower bounds are concave and all remaining constraints are convex, Problem~\eqref{prob_sca} is convex and can be efficiently solved using standard convex optimization solvers. After each iteration, the operating point is updated and the convex approximation is rebuilt until
$
|\eta^{(t)}-\eta^{(t-1)}|
\le
\epsilon
$,
where $\epsilon>0$ is the convergence tolerance. The iterative procedure terminates when the objective variation between two consecutive iterations becomes sufficiently small.

The overall procedure of the proposed framework is summarized in Algorithm~\ref{alg_greedy_sca}.

\begin{algorithm}[t]
\caption{Greedy-SCA Algorithm for PASS-Assisted RSMA}
\label{alg_greedy_sca}
\begin{algorithmic}[1]
\STATE Initialize $\mathcal A=\emptyset$.
\FOR{$m=1,\ldots,M$}
\FOR{each $n\in\mathcal N\setminus\mathcal A$}
\STATE Set $\mathcal A_n=\mathcal A\cup\{n\}$ and compute 
$\min_{k\in\mathcal K}|h_k(\mathcal A_n)|^2$.
\ENDFOR
\STATE Select
$
n^\star=
\arg\max_{n\in\mathcal N\setminus\mathcal A}
\min_{k\in\mathcal K}
|h_k(\mathcal A_n)|^2
$.
\STATE Update $\mathcal A\leftarrow\mathcal A\cup\{n^\star\}$.
\ENDFOR
\STATE Construct $\boldsymbol{\alpha}$ and compute $H_k=|h_k(\boldsymbol{\alpha})|^2$, $\forall k\in\mathcal K$.
\STATE Initialize feasible $\mathbf p^{(0)}$, $\mathbf C^{(0)}$, and $\eta^{(0)}$; set $t=0$.
\REPEAT
\STATE Construct $\widetilde R_{p,k}^{(t)}$ and $\widetilde R_{c,k}^{(t)}$.
\STATE Solve Problem~\eqref{prob_sca} to obtain $\mathbf p^{(t+1)}$, $\mathbf C^{(t+1)}$, and $\eta^{(t+1)}$.
\STATE Set $t\leftarrow t+1$.
\UNTIL{$|\eta^{(t)}-\eta^{(t-1)}|\le\epsilon$}.
\end{algorithmic}
\end{algorithm}

\subsection{Complexity Analysis}

The proposed framework consists of greedy PA activation followed by iterative SCA-based resource allocation. During the activation process, a total of $\sum_{m=0}^{M-1}(N-m)$ candidate evaluations are performed. Since each evaluation requires computing the channel metric for all $K$ users, the computational complexity of the greedy stage is $\mathcal{O}(KMN)$. For the selected PA configuration, each SCA iteration solves problem \eqref{prob_sca}, which involves $(2K+2)$ optimization variables. Using a standard interior-point solver, the computational complexity of each convex subproblem is approximately $\mathcal{O}((2K+2)^3)$. Therefore, after $T$ SCA iterations, the overall computational complexity is given by $\mathcal{O}(KMN)+\mathcal{O}\left(T(2K+2)^3\right)$. Compared with exhaustive activation, which requires evaluating all $\binom{N}{M}$ activation patterns, the proposed framework substantially reduces computational complexity.

\section{Numerical Results}
\label{sec:results}

In this section, numerical results are presented to evaluate the proposed PASS-assisted RSMA framework under imperfect SIC. The minimum user rate is adopted as the performance metric since it directly reflects the considered max--min fairness objective.
Unless otherwise specified, $N=20$ candidate PAs are available, among which $M=6$ PAs are activated. A total of $K=4$ users are uniformly distributed within the service region defined by $x_k\in[0,20]$ m and $y_k\in[2,10]$ m. The carrier frequency is set to $28$ GHz. The waveguide height, effective refractive index, and attenuation coefficient are set to $d=3$ m, $n_{\rm eff}=1.4$, and $\kappa_{\rm w}=0.01$ m$^{-1}$, respectively. The maximum transmit power and SIC imperfection coefficient are fixed at $P_{\max}=30$ dBm and $\rho=0.05$, respectively. The noise power is set to $-90$ dBm and the convergence tolerance is $\epsilon=10^{-4}$. All results are averaged over $500$ independent user-location realizations.

The proposed framework is compared with the following benchmark schemes:

\begin{itemize}

\item \textbf{PASS-RSMA with Fixed Activation}: A PASS-assisted RSMA system employing a predetermined PA activation pattern, where only common-rate and power allocation are optimized.

\item \textbf{PASS-RSMA with Random Activation}: The same RSMA optimization framework is employed, while the active PAs are randomly selected.

\item \textbf{PASS-NOMA}: PASS-assisted NOMA transmission with optimized power allocation under the same imperfect SIC model and max--min fairness objective.

\end{itemize}

\begin{figure*}[!t]
\centering

\subfloat[]{%
\includegraphics[width=0.24\textwidth,height=3.1cm]{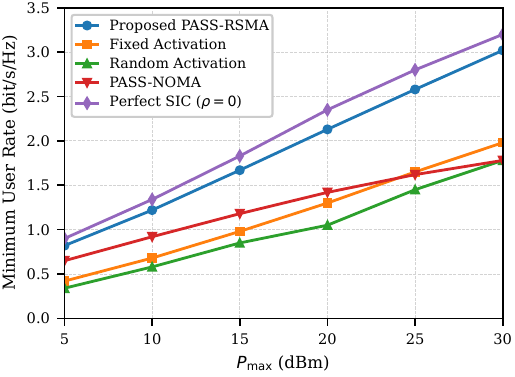}%
\label{fig_power}}
\hfil
\subfloat[]{%
\includegraphics[width=0.24\textwidth,height=3.1cm]{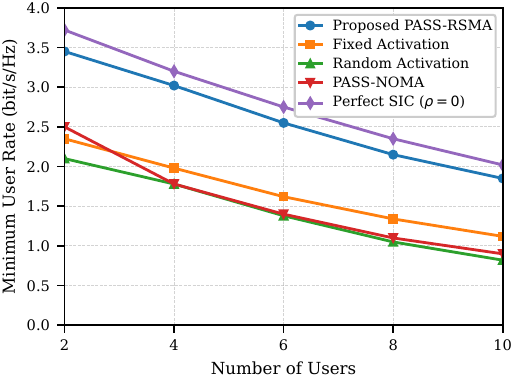}%
\label{fig_users}}
\hfil
\subfloat[]{%
\includegraphics[width=0.245\textwidth,height=3.18cm]{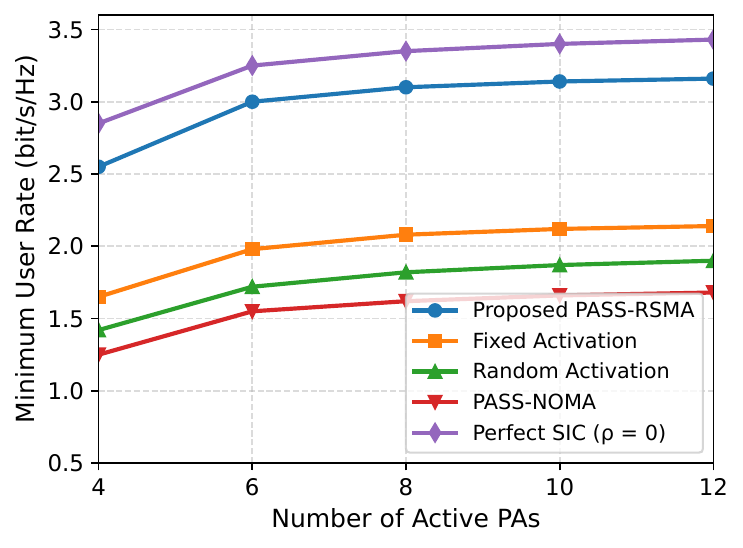}%
\label{fig_M}}
\hfil
\subfloat[]{%
\includegraphics[width=0.24\textwidth,height=3.1cm]{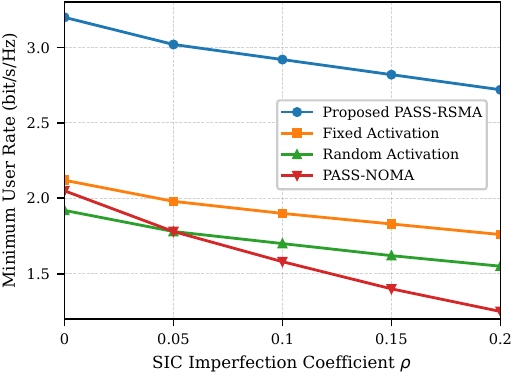}%
\label{fig_rho}}

\caption{Minimum user rate versus (a) maximum transmit power $P_{\max}$, (b) number of users $K$, (c) number of active PAs, and (d) SIC imperfection coefficient $\rho$.}
\label{fig_results}

\end{figure*}

Fig.~\ref{fig_power} depicts the minimum user rate versus the maximum transmit power $P_{\max}$. As expected, the minimum rate increases with the available transmit-power budget for all schemes. Nevertheless, the proposed PASS-RSMA framework consistently achieves the highest minimum user rate across the entire transmit-power range. The performance gain over the random-activation benchmark validates the effectiveness of the proposed channel-aware PA selection strategy. Moreover, the improvement relative to the fixed-activation benchmark highlights the benefit of adaptive PASS reconfiguration. Compared with PASS-NOMA, the proposed framework provides substantially higher fairness performance, demonstrating the advantage of RSMA-based interference management under imperfect SIC.  The performance gap also increases with the transmit-power budget, further highlighting the effectiveness of adaptive PA activation. 

Fig.~\ref{fig_users} shows the minimum user rate as a function of the number of users $K$. As $K$ increases, the available resources must be shared among a larger number of users, while the bottleneck-user effect becomes more pronounced, thereby reducing the achievable minimum rate. Despite the increased system load, the proposed framework consistently maintains the highest minimum user rate across the entire range of user densities. 

Fig.~\ref{fig_M} illustrates the average minimum user rate versus the number of activated PAs $M$. Increasing the number of active PAs initially improves the achievable minimum rate by enhancing the effective user-channel gains. However, the performance gain gradually saturates as more PAs are activated. This behavior stems from the coherent superposition of the radiated signals and the associated power normalization, which limit the incremental benefit of activating additional PAs. The proposed framework consistently outperforms the benchmark schemes across the entire range of $M$, demonstrating the effectiveness of channel-aware PA selection. These results highlight the importance of adaptive PA activation in PASS-assisted RSMA systems.

Fig.~\ref{fig_rho} evaluates the robustness of the considered schemes against SIC imperfections by varying the residual interference coefficient $\rho$. As $\rho$ increases, a larger fraction of the common-stream interference remains after SIC, thereby degrading private-stream decoding and reducing the achievable minimum rate. Nevertheless, the proposed PASS-RSMA framework exhibits the slowest performance degradation and consistently maintains the highest minimum rate across the considered SIC-imperfection range. These results confirm that the combination of adaptive PA activation and RSMA-based interference management effectively mitigates the adverse impact of imperfect SIC. Furthermore, the performance gap between the proposed framework and PASS-NOMA becomes more pronounced as $\rho$ increases, highlighting the robustness advantage of RSMA in the presence of residual interference.

\section{Conclusion}

This paper proposed a joint antenna activation and resource allocation framework for PASS-assisted RSMA systems under imperfect SIC. By explicitly incorporating residual SIC interference into the optimization process, a fairness-oriented max–min rate problem was formulated and efficiently addressed using a low-complexity greedy-SCA algorithm. The obtained results demonstrated that adaptive PASS reconfiguration significantly enhances the fairness performance of RSMA and provides improved resilience to SIC imperfections compared with conventional benchmark schemes. These findings highlight the potential of jointly exploiting PASS-enabled channel enhancement and RSMA-based interference management to improve fairness in practical beyond-5G and 6G wireless networks.

	\bibliographystyle{IEEEtran}
	\bibliography{Main_Document}

	\vfill
	
\end{document}